\documentclass[aps,twocolumn,showpacs,byrevtex,prl,reprint,nofootinbib]{revtex4-1}
\usepackage{xspace}

\usepackage{epsfig}
\usepackage{dcolumn}
\usepackage{bm}
\usepackage{color}
\usepackage{ulem}
\usepackage{pstricks}
\usepackage{pst-node}
\usepackage{rotating}
\usepackage{times}
\usepackage[english]{babel}
\addto{\captionsenglish}{%

}
\usepackage[colorlinks,linkcolor=blue,citecolor=blue]{hyperref}
\usepackage{lineno}

\newcommand{\EE}{e^+e^-}
\newcommand{\BB}{B\bar{B}}

\newcommand{\psp}{\psi(3686)}
\newcommand{\jpsi}{J/\psi}
\newcommand{\ar}{\rightarrow}
\newcommand{\llb}{\Lambda\bar{\Lambda}}

\newcommand{\bfg}{\begin{figure}}
\newcommand{\efg}{\end{figure}}
\newcommand{\bitm}{\begin{itemize}}
\newcommand{\eitm}{\end{itemize}}
\newcommand{\bnum}{\begin{enumerate}}
\newcommand{\enum}{\end{enumerate}}
\newcommand{\btbl}{\begin{table*}}
\newcommand{\etbl}{\end{table*}}
\newcommand{\btbu}{\begin{tabular}}
\newcommand{\etbu}{\end{tabular}}
\newcommand{\bcl}{\begin{center}}
\newcommand{\ecl}{\end{center}}
\newcommand{\bbt}{\bibitem}
\newcommand{\beq}{\begin{equation}}
\newcommand{\eeq}{\end{equation}}
\newcommand{\beqr}{\begin{eqnarray}}
\newcommand{\eeqr}{\end{eqnarray}}

\begin{document}
\normalsize
\parskip=5pt plus 1pt minus 1pt
\title{\boldmath
Measurement of the cross section for $\EE\ar\llb$ and evidence of the decay $\psi(3770)\ar\llb$
}
\author{
M.~Ablikim$^{1}$, M.~N.~Achasov$^{10,b}$, P.~Adlarson$^{67}$, S. ~Ahmed$^{15}$, M.~Albrecht$^{4}$, R.~Aliberti$^{28}$, A.~Amoroso$^{66A,66C}$, M.~R.~An$^{32}$, Q.~An$^{63,49}$, X.~H.~Bai$^{57}$
% ...
\\
\vspace{0.2cm}
(BESIII Collaboration)\\
\vspace{0.2cm} {\it
$^{1}$ Institute of High Energy Physics, Beijing 100049, People's Republic of China\\
% ...
$^{4}$ Bochum Ruhr-University, D-44780 Bochum, Germany\\
% ...
$^{10}$ G.I. Budker Institute of Nuclear Physics SB RAS (BINP), Novosibirsk 630090, Russia\\
% ...
$^{15}$ Helmholtz Institute Mainz, Staudinger Weg 18, D-55099 Mainz, Germany\\
% ...
$^{28}$ Johannes Gutenberg University of Mainz, Johann-Joachim-Becher-Weg 45, D-55099 Mainz, Germany\\
% ...
$^{32}$ Liaoning Normal University, Dalian 116029, People's Republic of China\\
% ...
$^{49}$ State Key Laboratory of Particle Detection and Electronics, Beijing 100049, Hefei 230026, People's Republic of China\\
% ...
$^{57}$ University of Jinan, Jinan 250022, People's Republic of China\\
% ...
$^{63}$ University of Science and Technology of China, Hefei 230026, People's Republic of China\\
% ...
$^{66}$ University of Turin and INFN, (A)University of Turin, I-10125, Turin, Italy; (B)University of Eastern Piedmont, I-15121, Alessandria, Italy; (C)INFN, I-10125, Turin, Italy\\
$^{67}$ Uppsala University, Box 516, SE-75120 Uppsala, Sweden\\
% ...
$^{b}$ Also at the Novosibirsk State University, Novosibirsk, 630090, Russia\\
% ...
}
}

\date{\today}

\begin{abstract}
The Born cross section
 of the process $\EE\ar\llb$ is measured at 33 center-of-mass energies between 3.51 and 4.60 GeV using data corresponding to the total integrated luminosity of 20.0 fb$^{-1}$ collected  with the BESIII detector at the BEPCII collider. Describing the energy dependence of the cross section requires
a contribution from the $\psi(3770)\ar\llb$ decay, which is fitted with a significance of 4.6$-$4.9$\sigma$ including the systematic uncertainty.
The lower bound on its branching fraction is $2.4\times10^{-6}$ at the 90\% confidence level (C.L.), 
at least an order of magnitude larger than expected from predictions using a scaling based on observed electronic widths. This result indicates the importance of effects from vector charmonium(-like) states when interpreting data in terms of {\it e.g.}, electromagnetic structure observables. 
The data do not allow for definite conclusions on the interplay with other vector charmonium(-like) states, and we set 90\% C.L.
 upper limits for the products of their electronic widths and the branching fractions.  
\end{abstract}
\maketitle

Two-body baryonic decays of vector ($J^{PC} = 1^{--}$) charmonium(-like) resonances provide a testing ground of predictions from 
quantum chromodynamics~\cite{Farrar,Briceno}.
The $\psi(3770)$ vector meson is believed to be a conventional $c\bar{c}$ state located above the open-charm threshold and is expected to decay into a $D\bar{D}$ meson pair with a branching fraction of at least 99\%~\cite{Eichten}. However, the decay modes to light quark systems would be considerably enhanced if the $\psi(3770)$ would include gluonic or light quark and antiquark constituents
~\cite{Close:2005iz}.
In 2003, the BES collaboration observed the first non-$D\bar{D}$  decay of 
$\psi(3770)$ into $\jpsi\pi^{+}\pi^{-}$~\cite{hepnp28_325, hepnp28_325-1}. 
Subsequently, the CLEO collaboration confirmed the observation and found more non-$D\bar{D}$ decays of 
$\psi(3770)$~\cite{prl96_082004,prl96_082004-1,prl96_082004-2}
and the first decay into light-quark hadrons $\psi(3770)\ar\phi\eta$~\cite{prd73_012002}. 

The production of light quark baryon-antibaryon ($\BB$) final states leads to relatively simple topologies.
In an early study, the BESIII collaboration found evidence for the interference effect in $\EE\ar p\bar{p}$ in the vicinity of $\psi(3770)$~\cite{Ablikim:2014jrz}. However, the data did not allow to uncover the mechanism of $\psi(3770)$ charmless decays.
Thus, the experimental study of $\EE\ar\BB$ will be a good search-ground for clarifying the nature of the charmless decays and even non-$D\bar{D}$ decays of $\psi(3770)$~\cite{Wan:2021vny, Cao:2021asd, Xia:2015mga}.

In the past two decades, several vector states were observed at energies between 3.7 and 4.7 GeV at various $\EE$ colliders. Four charmonium(-like) 
states predicted by potential models~\cite{Farrar} $\psi(3770)$, $\psi(4040)$, $\psi(4160)$ and $\psi(4415)$ have been observed as enhancements
in the inclusive hadronic cross section~\cite{PDG2016,Wang:2019ugd}. 
In addition, new states such as $Y(4230)$, $Y(4260)$, $Y(4360)$, $X(4390)$, and $Y(4660)$, were reported using the initial state radiation (ISR) processes $\EE\ar\gamma_{ISR}\pi^{+}\pi^{-}\jpsi(\psp)$ at the {\it BABAR}~\cite{BABAR01,BABAR01-1,BABAR01-2,BABAR01-3} and Belle~\cite{BELLE01,BELLE01-1,BELLE01-2,BELLE01-3,BELLE01-4} experiments, or in energy-scan experiments at the
CLEO-c~\cite{CLEO} and BESIII~\cite{BESIIIAB, BESIIIAB01, BESIIIAB03, BESIIIAB04, BESIIIAB05, BESIIIAB06} experiments.
Up to now, no evidence for decay modes into light-quark baryon--antibaryon pairs of these charmonium(-like) states has been
found. The overpopulation of vector charmonium(-like) resonances with respect to predictions from potential models,
and the difficulty in describing the properties of these states make them attractive candidates for exotic states~\cite{QCD}.
In addition, knowledge of the  vector charmonium(-like) coupling to the $\BB$ final states is crucial for understanding the electromagnetic structure of the baryons. In Refs.~\cite{Dobbs:2014ifa,Dobbs:2014ifa-1}, the timelike electromagnetic form factors for the ground-state octet baryons
were determined based on the CLEO-c data. 
They assumed that the branching fractions of $\psi(3770)$ to the $\BB$ final states scale with
the decay widths into a pair of electrons (electronic decay widths)
when comparing to the $\psi(3686)$ state, {\it e.g.}, one estimates a negligible branching fraction  ${\cal B}(\psi(3770)\to\Lambda\bar\Lambda)\approx 5\times10^{-7}$.

In this paper, we present a measurement of the Born cross section  for the $\EE\ar\llb$ process
using data corresponding to a total integrated luminosity of 20.0 fb$^{-1}$~\cite{ABC, Ablikim:2015nan, Ablikim:2015zaa}
collected at center-of-mass (c.m.) energies $\sqrt{s}$ between 3.51 and 4.60 GeV with the BESIII detector~\cite{Wang:2007tv, Wang:2007tv-1} at the BEPCII collider~\cite{BESIII}. We extract the $\Lambda$ effective form factor and report an evidence
 of the $\psi(3770)\ar\llb$ process by fitting the $\EE\ar\llb$ dressed cross section. 

Candidates for $\EE\ar\llb$ events are reconstructed using the  $\Lambda\ar p\pi^-$ and $\bar\Lambda\ar\bar{p}\pi^+$ decay modes. 
The detection efficiency is determined by Monte Carlo (MC) simulations. A sample of 100,000 events is simulated for each of the 33 c.m.\ 
energy points. The production
process is simulated by the \textsc{kkmc} generator~\cite{kkmc,kkmc-1} that includes corrections for ISR effects.
 The $\Lambda$ and $\bar\Lambda$ decays are handled by the \textsc{evtgen}~\cite{evt2,evt2-01} program. 
The response of the BESIII detector is modeled with MC simulations
using a framework based on \textsc{geant}{\footnotesize 4}~\cite{geant4, geant4-01}.

Tracks of charged particles are reconstructed in the multilayer drift chamber with a helical fit requiring a good quality~\cite{Asner:2008nq}.
These tracks should be within $|\cos\theta|<0.93$, where $\theta$ is the polar angle with respect to the $e^{+}$ beam direction. 
Events with two successfully reconstructed negatively charged and two positively charged particles
are kept for further analysis.

To reconstruct $\Lambda(\bar\Lambda)$ candidates,
we apply a secondary vertex fit~\cite{XUM} to all pairs of positive and negative charged particles. The corresponding $\chi^{2}$ value is required to be less than $500$. 
The track combination with minimum ${|M_{p\pi^{-}}-m_{\Lambda}|^{2} + |M_{\bar{p}\pi^{+}}-m_{\Lambda}|^{2}}$ is selected, where $M_{p\pi^{-}}(M_{\bar{p}\pi^{+}})$ is the invariant mass of the $p\pi^{-}(\bar{p}\pi^{+})$ pair, and $m_{\Lambda}$ is the world-average $\Lambda$ mass value from the Particle Data Group (PDG)~\cite{PDG2016}. 
To further suppress background from non-$\Lambda$ processes, 
the $\Lambda$ decay length is required to be larger than zero, where
the observed negative decay lengths are caused by the limited detector resolution.  
Here the misreconstruction ratio for $\Lambda$ particle is found to be less than 1\% based on the study of MC simulation.

\begin{figure}[!htbp]
\includegraphics[width=0.45\textwidth]{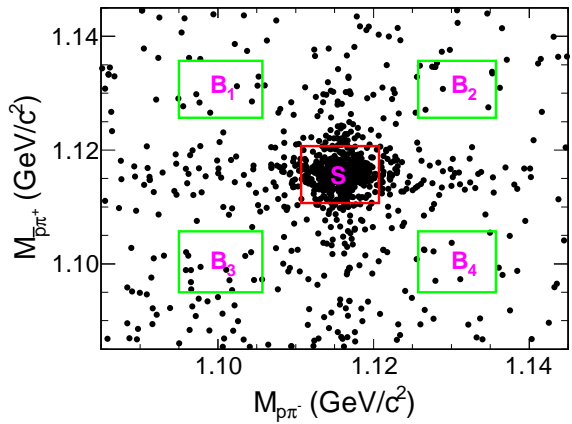}
\caption{Distribution of $M_{p\pi^{-}}$ versus $M_{\bar{p}\pi^{+}}$ of the accepted candidates from all data samples, where the red box shows the signal region, the green boxes denote the selected sideband regions.
}
\label{scatter_plot::llb:mc}
\end{figure}

To further suppress the background and to improve the mass resolution, a four-constraint (4C) kinematic fit imposing energy-momentum conservation
is applied for the $\llb$ hypothesis. The $\chi^{2}_{4C}$ of the fit is required to be less than 200, which can improve 
the resolutions of signals significantly in addition to suppressing backgrounds for a soft $\pi^{0}$ and radiated photons event.
Figure~\ref{scatter_plot::llb:mc} shows the distribution of
$M_{p\pi^{-}}$ versus $M_{\bar{p}\pi^{+}}$ of the accepted candidates from all data samples.
Clear peaks around the $\Lambda$ known mass can be discerned.
The invariant mass $M_{p\pi^{-}}$ is required to be within 5 MeV/$c^{2}$ of the known
$\Lambda(\bar\Lambda)$ mass (signal region marked by $S$).
After applying the above selection criteria, the survived background events
 are mainly from non-$\Lambda(\bar\Lambda)$ events, such as $\EE\ar\pi^{+}\pi^{-}p\bar{p}$.
The background yield in the signal region is estimated
using four sideband regions $B_{i}$, where $i=1,2,3,4$, each with the same area as the signal region. The  regions are shown in Fig.~\ref{scatter_plot::llb:mc}, and
the exact ranges are given in the Supplemental Material~\cite{ABC}.
The signal yield $N_{obs}$ for $\EE\ar\llb$ events at each energy point 
can then be extracted by subtracting the number of events in the sideband regions from the number of events in the signal region, $N_S$: $N_{obs} = N_{S} - \frac{1}{4}\sum^{4}_{i=1}N_{B_{i}}$, and they are listed in Table~\ref{tab:signal:yields:DD}.

The ISR corrected (``dressed'') cross section
 $\sigma^{\rm dr}$ for the process $\EE\ar\llb$ is defined as
\begin{equation}
\sigma^{\rm dr}(s) =\frac{N_{obs}}{{\cal{L}}(1 + \delta)\epsilon\ {\cal B}^{2}(\Lambda\ar p\pi^{-})}\ ,
\end{equation}
where ${\cal{L}}$ is the integrated luminosity at given c.m.\  energy $\sqrt{s}$, $(1
+ \delta)$ is the ISR correction factor~\cite{Jadach:2000ir,kkmc-1}, $\epsilon$ is the detection
efficiency, and the branching fraction ${\cal B}(\Lambda\ar p\pi^{-})=(63.9\pm0.5)\%$ is  taken from PDG.  The ISR correction factor is obtained using the calculation described in Ref.~\cite{Kuraev:1985hb}, where the dressed cross sections are adopted as initial input and are iterated to obtain stable result.
The dressed cross section is related to the Born cross section via the vacuum polarization factor $\frac{1}{|1 - \Pi|^2}$~\cite{Actis:2010gg} as $\sigma^{\rm dr} = \ \sigma^{B}/|1 - \Pi|^{2}$ (further details are provided in the Supplemental Material~\cite{ABC}). 

Systematic uncertainties on the cross section measurement 
mainly come from the luminosity measurement, the $\Lambda$ reconstruction, the 4C kinematic fit, the branching fraction for the decay $\Lambda\ar p\pi^{-}$, the line-shape description, and the physical model dependence. 
The uncertainty due to the vacuum polarization is negligible. The integrated luminosity is measured by $\EE\ar(\gamma)\EE$ events with a similar method to Ref.~\cite{Ablikim:2015nan} with an uncertainty of 1.0\%.
The systematic uncertainty of the $\Lambda(\bar\Lambda)$ reconstruction incorporating the tracking, the mass window of $\Lambda(\bar\Lambda)$, and the decay length of $\Lambda(\bar\Lambda)$
is studied using a control sample of $\psp\ar\llb$ decay ($\sim$ 20000 events) with the same method as introduced in
Refs.~\cite{Ablikim:2016iym, Ablikim:2016iym-1,Ablikim:2016iym-2,Ablikim:2016iym-3, Ablikim:2016iym-4, Ablikim:2016iym-5, BESIII:2021gca}. The signal MC sample
is simulated using a DIY model~\cite{evt2} implementing the joint angular distribution from Refs.~\cite{Faldt:2017kgy,nature}.
The efficiency difference between data and MC simulation is found to be 0.5\% for the $\Lambda$ reconstruction and 1.5\% for the $\bar\Lambda$ reconstruction. 
The uncertainty from the 4C kinematic fit is
studied using the control sample of $\psp\ar\llb$ decays with and without performing a 4C kinematic fit. The relative change of 1.0\% is assigned as the systematic uncertainty.
The uncertainty of the branching fraction for $\Lambda\ar p\pi^{-}$ from the PDG~\cite{PDG2016} is 0.8\%, and is propagated to the final result.
The uncertainty from the line-shape description is estimated with an alternative input cross section line shape 
based on a simple power-law function. The change of the efficiency, 2.6\%, is taken as the systematic uncertainty.
The uncertainty due to the physical model dependence is estimated to be 2.5\% by comparing the efficiencies between phase space and the DIY model incorporating the $\Lambda$ transverse polarization and spin correlation based on the control sample of $\psi(3770)\ar\llb$ decays.
 Assuming all sources are independent,
the total systematic uncertainty on the cross section measurement 
is determined to be 4.3\% by adding these sources in quadrature. The correlations
 for the different points are negligible due to the limited statistics.

\begin{figure}[!htbp]
\includegraphics[width=0.45\textwidth]{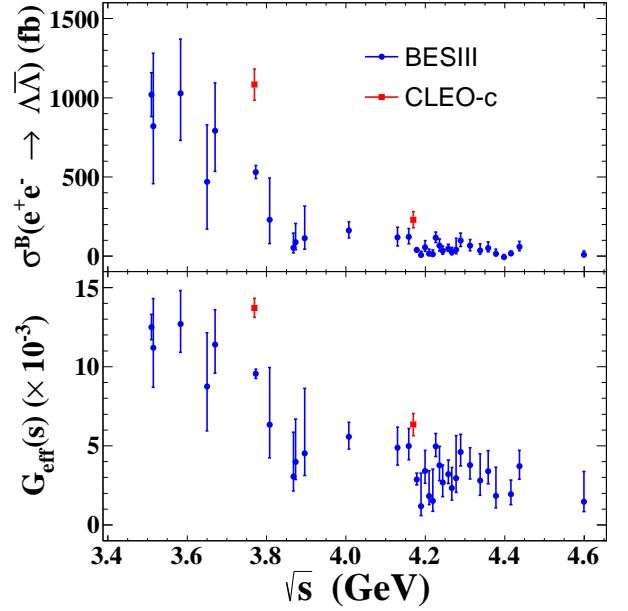}\\
\caption{The measured Born cross section (top) and $\Lambda$ effective form factor (bottom) for $\EE\ar\llb$ as a function of the c.m. energy, where the uncertainties include the statistical and systematic ones.}
\label{Cross_SC_com}
\end{figure}

The extracted Born cross sections at each energy point are listed in Table~\ref{tab:signal:yields:DD} and shown in Fig.~\ref{Cross_SC_com}(top)
together with the CLEO-c results at 3.770 and 4.160 GeV~\cite{Dobbs:2014ifa,Dobbs:2014ifa-1}.
Our results are significantly lower than
those of CLEO-c at both energy points.
Figure~\ref{Cross_SC_com} (bottom)
shows the extracted energy dependence of the $\Lambda$ effective form factor $G_{\rm eff}(s)$ defined as~\cite{Ablikim:2019kkp}
\begin{equation}\label{EFF}
G_{\rm eff}(s)  = \sqrt{\frac{3s\tau\sigma^{B}}{2\pi\alpha^2\beta(2\tau+1)}}\ ,
\end{equation}
where
$\alpha$ is the fine-structure constant, $\beta =\sqrt{(\tau-1)/\tau}$ is the $\Lambda$ velocity and  $\tau = {s}/({4m^{2}_{\Lambda}})$.
\begin{table}[!hpt]
\begin{center}
\caption{\small The signal yield
$N_{obs}$ and Born cross sections $\sigma^{B}$ obtained at the 33 c.m.\ energy points. The values in the brackets represent the upper limit at 90\% C.L. calculated with a profile likelihood method~\cite{Lundberg:2009iu} taking into account the systematic uncertainty.
The first and second uncertainties for $\sigma^{B}$ are statistical and systematic, respectively.
}
 \begin{tabular}{c r@{.}l r@{}l@{$\pm$~}l@{~}l} \hline\hline
 $\sqrt{s}$ (GeV) &\multicolumn{2}{c}{$N_{obs}$} &\multicolumn{4}{c}{$\sigma^{B}$ (fb)} \\ \hline
3.5100  &61   &$0^{+7.8}_{-7.8}$              &1020 &$^{+130}_{-130}$  &44 &             \\
3.5146  &5    &$0^{+2.8}_{-2.2}$  ($<$ 9.7)   &820  &$^{+460}_{-360}$  &35 &($<$ 1600)   \\
3.5815  &13   &$0^{+4.3}_{-3.7}$              &1030 &$^{+340}_{-290}$  &44 &             \\
3.6500  &3    &$0^{+2.3}_{-1.9}$  ($<$ 6.8)   &470  &$^{+360}_{-230}$  &20 &  ($<$ 1000) \\
3.6702  &10   &$0^{+3.8}_{-3.2}$              &790  &$^{+370}_{-250}$  &34 &             \\
3.7730  &261  &$0^{+16.2}_{-16.2}$            &530  &$^{+33}_{-33}$    &22 &             \\
3.8077  &2    &$0^{+2.3}_{-1.3}$  ($<$ 5.3)   &230  &$^{+260}_{-150}$  &10 &($<$ 610)    \\
3.8675  &1    &$0^{+1.8}_{-0.6}$  ($<$ 3.7)   &52   &$^{+94}_{-31}$    &2  &($<$ 190)    \\
3.8715  &1    &$7^{+2.3}_{-0.6}$  ($<$ 5.3)   &88   &$^{+120}_{-47}$   &4  &($<$ 270)    \\
3.8962  &1    &$0^{+1.8}_{-0.6}$  ($<$ 3.7)   &110  &$^{+200}_{-68}$   &5  &($<$ 420)    \\
4.0076  &13   &$0^{+4.3}_{-3.7}$              &160  &$^{+54}_{-46}$    &7  &             \\
4.1301  &6    &$0^{+3.3}_{-2.7}$  ($<$ 10.0)   &120  &$^{+65}_{-53}$    &5  &($<$ 200)    \\
4.1585  &7    &$7^{+3.3}_{-2.7}$  ($<$ 13.7)  &120  &$^{+52}_{-43}$    &5  &($<$ 220)    \\
4.1783  &18   &$0^{+5.3}_{-4.2}$              &40   &$^{+12}_{-9}$     &2  &             \\
4.1893  &0    &$5^{+1.8}_{-0.5}$  ($<$ 3.7)   &7    &$^{+24}_{-7}$     &1  &($<$ 50)     \\
4.1996  &3    &$7^{+2.8}_{-1.7}$  ($<$ 8.3)   &56   &$^{+42}_{-26}$    &2  &($<$ 130)    \\
4.2097  &1    &$0^{+1.8}_{-0.6}$  ($<$ 3.7)   &16   &$^{+29}_{-10}$    &1  &($<$ 59)      \\
4.2188  &0    &$7^{+1.8}_{-0.6}$  ($<$ 3.7)   &11   &$^{+29}_{-10}$    &1  &($<$ 59)     \\
4.2263  &16   &$7^{+4.4}_{-3.8}$              &120  &$^{+32}_{-30}$    &5  &             \\
4.2358  &4    &$5^{+2.8}_{-2.3}$  ($<$ 9.7)   &66   &$^{+41}_{-34}$    &3  &($<$ 140)    \\
4.2439  &2    &$7^{+2.3}_{-1.8}$  ($<$ 6.8)   &34   &$^{+29}_{-23}$    &2  &($<$ 85)     \\
4.2580  &6    &$0^{+3.3}_{-2.2}$  ($<$ 11.0)  &48   &$^{+26}_{-18}$    &2  &($<$ 88)     \\
4.2669  &2    &$0^{+2.3}_{-1.3}$  ($<$ 5.3)   &25   &$^{+29}_{-16}$    &1  &($<$ 66)     \\
4.2778  &1    &$0^{+1.8}_{-0.6}$  ($<$ 3.7)   &40   &$^{+72}_{-24}$    &2  &($<$ 150)    \\
4.2889  &7    &$0^{+3.3}_{-2.7}$  ($<$ 12.4)  &98   &$^{+46}_{-38}$    &4  &($<$ 170)    \\
4.3128  &4    &$7^{+2.8}_{-2.2}$  ($<$ 9.7)   &65   &$^{+39}_{-31}$    &3  &($<$ 130)    \\
4.3379  &2    &$7^{+2.3}_{-1.8}$  ($<$ 6.8)   &35   &$^{+43}_{-24}$    &2  &($<$ 89)     \\
4.3583  &3    &$7^{+2.8}_{-1.7}$  ($<$ 8.3)   &51   &$^{+39}_{-24}$    &2  &($<$ 110)     \\
4.3776  &1    &$2^{+2.3}_{-1.0}$  ($<$ 5.3)   &15   &$^{+29}_{-13}$    &1  &($<$ 67)     \\
4.3980  &0    &{$0^{+0.8}_{-0.0}$}  ($<$ 2.0)  &0 &$^{+11}_{-0}$     &1  &($<$ 27)     \\
4.4156  &2    &$5^{+2.3}_{-1.7}$  ($<$ 6.8)   &16   &$^{+15}_{-11}$    &1  &($<$ 45)     \\
4.4370  &5    &$0^{+2.8}_{-2.2}$  ($<$ 9.7)   &59   &$^{+33}_{-26}$    &3  &($<$ 120)    \\
4.5995  &0    &$7^{+1.8}_{-0.8}$  ($<$ 3.7)   &9    &$^{+23}_{-8}$    &1  &($<$ 47)     \\
\hline\hline
\end{tabular}
\label{tab:signal:yields:DD}
\end{center}
\end{table}

\begin{figure}[!htbp]
\includegraphics[width=0.45\textwidth]{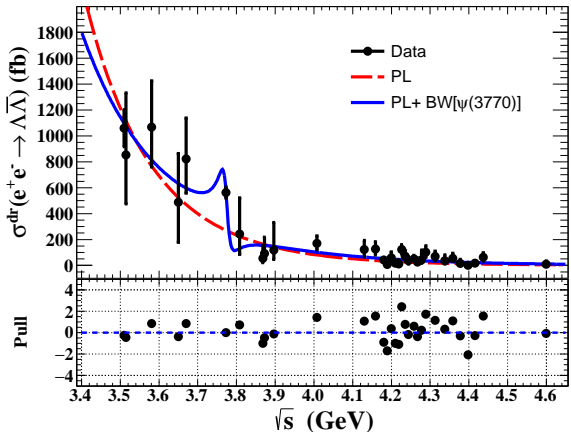}
\caption{The dressed cross section of the process $\EE\ar\llb$
is represented by the dots with error bars that include
statistical and systematic uncertainties. 
The red dashed line represents the fit with the power-law function only, while the solid blue line is  for the fit with the power-law function and the $\psi(3770)$ resonance.
The bottom panel shows the pull distribution for the fit with the resonance.
}
\label{lineshape_CS_4BW}
\end{figure}

\begin{figure}[!htbp]
\includegraphics[width=0.46\textwidth]{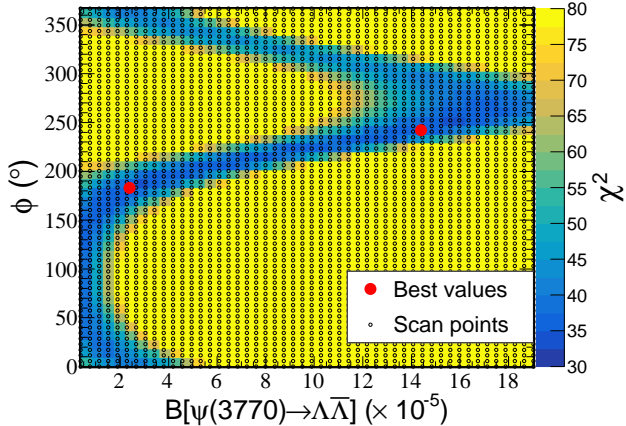}
\caption{The contour of ${\cal{B}}$ and $\phi$ on the distribution of $\chi^{2}$ values. The black open circles are the scan points for each set of parameters.The red points represent the center values of nominal values from the best fit. The x-axis starts from $0.1 \times 10^{-5}$.
}
\label{lineshape_CS_4BW:contour}
\end{figure}

The dressed cross section for the continuum
$\EE\ar\llb$ process is expected to have an asymptotic power-law behavior $\propto s^{-n}$ with the exponent $n\approx10$~\cite{Ablikim:2019kkp, Lepage:1980fj,Lepage:1980fj-1}.
A least-$\chi^{2}$ fit including statistical and systematic uncertainties  
to the power-law distribution describes the data points reasonably well, as shown with the dashed line in Fig.~\ref{lineshape_CS_4BW}. The fitted value of the exponent $n$ is not close to 10 within the uncertainty of 1$\sigma$,
 as shown in the column ``Fit I'' in Table~\ref{tab:Lam::RP}.
A fit with the coherent sum of the power-law function and a Breit-Wigner (BW) function
\begin{equation}
    \sigma^{\rm dr }({s})=\left|\sqrt{\sigma_{\rm 0}}{\left(\frac{M}{\sqrt{s}}\right)^{n}} + e^{i\phi}{\rm BW}({s})\right|^{2}
\end{equation}
is applied, where $M$ is the $\psi(3770)$ mass, 
$\sigma_{0}$ is the value of the continuum cross section at  $\psi(3770)$ and $\phi$ is the relative phase between the continuum and the resonance.
The BW function is
\begin{equation}
    {\rm BW}({s}) =\frac{\sqrt{\sigma_\psi}M\Gamma}{s-M^{2}+iM\Gamma}\  {\rm with}\ 
    \sigma_\psi\!=\!\frac{{12\pi (\hbar c)^2 {\Gamma}_{ee}{\cal{B}}}}{\Gamma M^2}\ ,
\end{equation} 
where $\Gamma$ and ${\Gamma}_{ee}$ are  the total and the electronic width of the $\psi(3770)$ resonance, respectively,
and ${\cal{B}}$ denotes the branching fraction to $\llb$. The solid line in Fig.~\ref{lineshape_CS_4BW} and the column "Fit II" in Table~\ref{tab:Lam::RP} shows the result of the fit with two solutions, where the mass and width of $\psi(3770)$ are fixed according to the PDG values~\cite{PDG2016}, and 
 $\sigma_0$, $n$, $\phi$ and $\sigma_\psi$ parameters are free. The improvement of the $\chi^2$ value gives a significance of 4.6$-$4.9$\sigma$ for the hypothesis with the $\psi(3770)$ resonance including the systematic uncertainty. The correlation coefficient between the resonance cross section $\sigma_\psi$ and the phase $\phi$ is almost equal to one. In the Fit II, two solutions 
are expected according to  
mathematical calculation~\cite{Zhu:2011ha, Zhu:2011ha-1}, 
but fits give the consistent results within the uncertainty of 1$\sigma$ due to statistics limitation. 
Figure~\ref{lineshape_CS_4BW:contour} shows the contour of ${\cal{B}}$ and $\phi$ on the distribution of $\chi^{2}$ values for each set of parameters.
Our results can be summarized by giving 90\% C.L. intervals  
$24<\sigma_\psi<1800$ fb and $2.4\times10^{-6}<{\cal B}<1.8\times10^{-4}$.
This represents the first evidence of the decay $\psi(3770)\ar\llb$.
This result is larger by  at least an order of magnitude than the prediction based on a scaling from the electronic branching
fraction value. This implies, that the $\psi(3770)$ resonance needs to
be considered when interpreting the CLEO-c data.
Note that the systematic uncertainties due to beam energy, mass and width of the $\psi(3770)$ resonance have been considered by varying the known value within one standard deviation, and they turn out to be negligible.

\begin{table}[!hbpt]
\begin{center}
\caption{ 
 Results of the fit to the dressed cross section for the $\EE\to\llb$ process, where two solutions in "Fit II" are provided.
The fitting procedure includes both statistical and systematic uncertainties except for the c.m.\ energy calibration.
 ${\cal{B}}$ is the branching fraction of the decay $\psi(3770)\ar\llb$ measured assuming  ${\cal{B}}_{ee}=9.7\times10^{-6}$, the central value of the world average~\cite{PDG2016}, and
 $\Gamma_{ee}=\Gamma_{\psi(3770)}{\cal{B}}_{ee} = (261.2 \pm 21.3)$ eV.  }
\begin{tabular}{lr@{ $\pm$ }lr@{ $\pm$ }lr@{ $\pm$ }l}  \hline\hline
    &\multicolumn{2}{c}{Fit I}   &\multicolumn{4}{c}{Fit II}
   \\ \hline
$\sigma_{0}$ (fb) &379 &22&\multicolumn{4}{c}{320  $^{+750}_{-340}$} \\
$n$ &8.8&0.4&\multicolumn{4}{c}{8.2 $\pm$ 0.6} \\
$\phi$ ($^{\circ}$)&\multicolumn{2}{c}{--}&\multicolumn{2}{c}{183$^{+57}_{-40}$} &\multicolumn{2}{c}{240$^{+17}_{-115}$}\\
$\sigma_{\psi}$  (fb)&\multicolumn{2}{c}{0(fixed)}&\multicolumn{2}{c}{240$^{+1470}_{-190}$}  &\multicolumn{2}{c}{1440$^{+270}_{-1390}$} \\
$\chi^{2}/ndof$&\multicolumn{2}{c}{62.0/31} 
 &\multicolumn{4}{c}{34.6/29}  \\ \hline
${\cal{B}}$ $(\times 10^{-5})$ &\multicolumn{2}{c}{--}&\multicolumn{2}{c}{2.4$^{+15.0}_{-1.9}$} &\multicolumn{2}{c}{14.4$^{+2.7}_{-14.0}$}\\
\hline\hline
\end{tabular}
\label{tab:Lam::RP}
\end{center}
\end{table}
Finally, we have included an additional charmonium(-like) state (i.e. $\psi(4040)$, $\psi(4160)$, $Y(4260)$, or $\psi(4415)$) in the fit, one at a time.
 It turns out that the exponent and significance for the $\psi(3770)$ state are consistent with 
a single resonance assumption.
Since the significance of each mentioned state is smaller than 3$\sigma$, we quote upper limits at the 90\% C.L. for the $\Gamma_{ee}{\cal B}$ products: 
$<5.5\times10^{-3}$ eV  for $\psi(4040)$, $<0.7\times10^{-3}$ eV for $\psi(4160)$, $<0.8\times10^{-3}$ eV for $Y(4260)$ and $<1.8\times10^{-3}$ eV for $\psi(4415)$ including the systematic uncertainty.
These results provide 
important information to understand the nature of charmonium(-like) states above the open charm threshold. In particular this concerns
their coupling to the $\BB$ final states and insight into the puzzle of large non-$D\bar{D}$ component of 
$\psi(3770)$.
\section{Acknowledgement}
\label{sec:acknowledgement}
The BESIII collaboration thanks the staff of BEPCII and the IHEP computing center for their strong support. This work is supported in part by National Key Research and Development Program of China under Contracts No. 2020YFA0406400 and No. 2020YFA0406300; National Natural Science Foundation of China (NSFC) under Contracts No. 11625523, No. 11635010, No. 11735014, No. 11822506, No. 11835012, No. 11875115, No. 11905236, No. 11935015, No. 11935016, No. 11935018, No. 11961141012, No. 12022510, No. 12035009, No. 12035013, No. 12047501, No. 12075107, and No. 12061131003; the Chinese Academy of Sciences (CAS) Large-Scale Scientific Facility Program; Joint Large-Scale Scientific Facility Funds of the NSFC and CAS under Contracts No. U1732263, and No. U1832207; CAS Key Research Program of Frontier Sciences under Contract No. QYZDJ-SSW-SLH040;
Fundamental Research Funds for the Central Universities under Grant No. lzujbky-2021-sp24;
100 Talents Program of CAS; INPAC and Shanghai Key Laboratory for Particle Physics and Cosmology; ERC under Contract No. 758462; European Union Horizon 2020 research and innovation program under Contract No. Marie Sklodowska-Curie Grant Agreement No. 894790; German Research Foundation DFG under Contract No. 443159800, Collaborative Research Center CRC 1044, FOR 2359, FOR 2359, GRK 214; Istituto Nazionale di Fisica Nucleare, Italy; Ministry of Development of Turkey under Contract No. DPT2006K-120470; National Science and Technology fund; Olle Engkvist Foundation under Contract No. 200-0605; STFC (United Kingdom); The Knut and Alice Wallenberg Foundation (Sweden) under Contract No. 2016.0157; The Royal Society, UK under Contracts No. DH140054, and No. DH160214; The Swedish Research Council; U. S. Department of Energy under Contracts No. DE-FG02-05ER41374, No. DE-SC-0012069.

\end{document}